\documentclass[12pt]{article}
\usepackage{amsmath,amssymb,amsthm}
\usepackage{mathbbol,bbm}
\usepackage[final]{showkeys}
\usepackage{bm}
\usepackage{calc}

\begin{document}

\begin{center}
{\LARGE Remarks on the violation of Bell's inequality in Josephson phase qubits} \\[18pt]
R.~Alicki\\[6pt]
Institute of Theoretical Physics and Astrophysics \\
University of Gda\'nsk, Poland \\[6pt]
\end{center}

\begin{abstract}
The consistency test and error estimation for the data concerning recently reported violation of
Bell inequality for Josephson phase qubits are presented in details. It is pointed out that the deviation of the Bell signal from the classical limit might still be challenged. 
\end{abstract}

In the recent Letter to Nature \cite{Nature} the authors reported the experimental results concerning violation of Clauser-Horn-Shimony-Holt (CHSH) inequality in the so-called Josephson phase qubits. They measured "a Bell signal of $ 2.0732\pm 0.0003$, exceeding the maximum amplitude of 2 for a classical system by 244 standard deviations". In this note I present a general analysis of a data's consistency and  errors valid for any experimental setting concerning violation of CHSH inequality. This led to the discovery of a typographical error in the data published in the Supplementary Information \cite{Sup},  corrected in the errata \cite{Err}. The main conclusion of this analysis is that the estimated error due to crosstalk between two Josephson junctions is higher than the initially reported $0.5\%$ but should be at least of the order of $1.5\%$ or even more if one takes into account that the optimization procedure toward the maximal Bell signal can maximize the crosstalk errors as well. Therefore, the deviation of the Bell signal from the classical limit, which is about $3.5\%$ are still comparable to the worst case errors. 
\par
Remind that in the experimental setting always two dichotomic observables (with values $\pm 1$) $Z_A, Z_B$  are measured for the system $A$ and $B$ respectively. Firstly, the same joint initial state $\rho$ is prepared for each run of the experiment. Then choosing a setting $(x,y)$ where $x= a,a'$ corresponds to $A$ and $y= b,b'$
to $B$ one applies two unitary gates $U_x$ and $U_y$ acting on $A$ and $B$ respectively to obtain an \emph{ideal density matrix } $\rho^0_{xy}= U_x U_y\rho U_y^{\dagger}U_x^{\dagger}$ in the absence of \emph{crosstalk} between $A$ and $B$ 
\footnote{Equivalently, one can use instead of four gates  $U_x ,U_y$, four observables $Z^x_A =U_x^{\dagger} Z_A U_x , Z^y_B = U_y^{\dagger}Z_B U_y$.}. However, the true density matrix after performing the gates $\rho^1_{xy}$ differs from the ideal one due to the cumulative effect of various error and noise sources. The measured and ideal Bell signals
$S^1$ and $S^0$ read, respectively  $(i=0,1)$
\begin{equation}
S^i = \mathrm{Tr}(\rho^i_{ab}Z_A Z_B) +\mathrm{Tr}(\rho^i_{a'b}Z_A Z_B)-\mathrm{Tr}(\rho^i_{ab'}Z_A Z_B)+\mathrm{Tr}(\rho^i_{a'b'}Z_A Z_B).
\label{Bsig}
\end{equation}
Denoting by $\eta_{xy} = \|\rho^1_{xy}-\rho^0_{xy}\|_{\mathrm{tr}}$ one obtains the following
estimation 
\footnote{The symbols $\|\cdot\|_{\mathrm{tr}}, \|\cdot\|_{\infty}$ denote trace norm and operator norm, respectively. The inequality $|\mathrm{Tr}(XY)|\leq\|X\|_{\mathrm{tr}} \|Y\|_{\infty}$ and the fact $\|Z_A\|_{\infty}=\|Z_B\|_{\infty}=1$ are used.}
\begin{equation}
|S^1 - S^0|\leq \sum_{xy} \eta_{xy} 
\label{maxer}
\end{equation}
and hence a corrected classical bound in terms of \emph{maximal errors} $\eta_{xy}$
\begin{equation}
|S^1|\leq 2 + \sum_{xy} \eta_{xy} .
\label{clb}
\end{equation}
The experimental data allows to compute for any setting $(x,y)$ the averages $\mathrm{Tr}(\rho^1_{xy}Z_{A(B)} )$ and hence the \emph{crosstalk parameters}
\begin{equation}
\delta_x = |\mathrm{Tr}(\rho^1_{xb}Z_A )-\mathrm{Tr}(\rho^1_{xb'}Z_A )|\ ,\  \delta_y = |\mathrm{Tr}(\rho^1_{ay}Z_B )-\mathrm{Tr}(\rho^1_{a'y}Z_B )| .
\label{clb1}
\end{equation}
Taking into account that in the absence of crosstalk $\mathrm{Tr}(\rho^0_{xb}Z_A )=\mathrm{Tr}(\rho^0_{xb'}Z_A) $ and $\mathrm{Tr}(\rho^0_{ay}Z_B )=\mathrm{Tr}(\rho^0_{a'y}Z_B )$ one obtains the following estimation
\begin{equation}
\delta_x = |\mathrm{Tr}[(\rho^1_{xb}-\rho^0_{xb})Z_A ]-\mathrm{Tr}[(\rho^1_{xb'}-\rho^0_{xb'})Z_A ]|\leq \eta_{xb}+\eta_{xb'}
\label{cross}
\end{equation}
and similarly
\begin{equation}
\delta_y \leq \eta_{ay}+\eta_{a'y}\ .
\label{cross1}
\end{equation}
In the case of a random choice of settings one can expect that the terms $\mathrm{Tr}[(\rho^1_{xb}-\rho^0_{xb})Z_A$ and $\mathrm{Tr}[(\rho^1_{xb'}-\rho^0_{xb'})Z_A ]$ in (\ref{cross}) may add or cancel equally likely. This leads to the 
rough estimation $\delta_x \simeq (\eta_{xb}+\eta_{xb'})/2$, etc., and hence
\begin{equation}
\eta\equiv\sum_{xy}\eta_{xy}\simeq \sum_{\mu=a,a',b,b'}\delta_{\mu}\equiv\delta\ .
\label{est}
\end{equation}
Using the corrected data of \cite{Err} one can compute 
\begin{equation}
\delta_a =0.0127,\delta_{a'}= 0.0176 , \delta_b= 0.0000 , \delta_{b'}= 0.0002 ,\delta= 0.0305.
\label{delta}
\end{equation}
Assuming the estimation (\ref{est}) for  the total error $\eta\simeq\delta$ one can notice that the reported violation of the classical limit is over two times larger than the estimated error.
However, this estimation might be too optimistic because the results were obtained using optimization procedure with respect to the settings (measurement angles). In particular, one can see the strong asymmetry between the crosstalk parameters $\delta_x$ and $\delta_y$. This suggests that the optimization procedure minimized the values of $\delta_b , \delta_{b'}$ making
the estimation (\ref{est}) less convincing and hence the main message of the paper \cite{Nature} might still be challenged.
Perhaps, it would be useful to compare the test based on Bell signal with a single system test proposed in \cite{A} and demonstrated in the case of a single photon polarisation in \cite{Gen}. In this approach the basic experimental limitation is a purity of an initial single qubit state which should be higher than $95\%$.


\end{document}